\documentclass[preprint,aps,nofootinbib,tightenlines]{revtex4}

\usepackage{graphicx}

\usepackage{amsmath}
\usepackage{amsfonts}
\usepackage{amssymb}

\newcommand{\be}{\begin{equation}}
\newcommand{\ee}{\end{equation}}
\newcommand{\bear}{\begin{eqnarray}}
\newcommand{\eear}{\end{eqnarray}}
\newcommand{\beqstar}{\begin{eqnarray*}}
\newcommand{\eeqstar}{\end{eqnarray*}}

\begin{document}

\preprint{HUTP-03/A051}

\vspace*{1.5cm} 
\title{TeV Symmetry and the Little Hierarchy Problem\vspace{0.5cm}}

\author{Hsin-Chia Cheng and Ian Low}
\affiliation{Jefferson Physical Laboratory,
Harvard University, Cambridge, MA 02138
\vspace*{0.5cm}}

\begin{abstract}
\vspace*{0.5cm} 

Constraints from precision electroweak measurements reveal no evidence
for new physics up to 5 -- 7 TeV, whereas naturalness requires new
particles at around 1 TeV to address the stability of the electroweak
scale. We show that this ``little hierarchy problem'' can be cured by
introducing a symmetry for new particles at the TeV scale. As an
example, we construct a little Higgs model with this new symmetry,
dubbed $T$-parity, which naturally solves the little hierarchy problem
and, at the same time, stabilize the electroweak scale up to 10
TeV. The model has many important phenomenological consequences,
including consistency with the precision data without any fine-tuning,
a stable weakly-interacting particle as the dark matter candidate, as
well as collider signals completely different from existing little
Higgs models, but rather similar to the supersymmetric theories with
conserved $R$-parity.

\end{abstract}


\maketitle

\section{Introduction}
\label{sec:introduction}

Standard Model is very successful in describing all known phenomena in
particle physics to date. It is nonetheless theoretically incomplete
as the mass-squared parameter for the Higgs doublet receives
quadratically divergent corrections at the quantum level and hence is
very sensitive to ultraviolate  physics. In order for the Higgs mass to
be naturally in the ${\cal O}(100)$ GeV range, new physics which
couples to the Higgs sector should appear at the scale $\sim$ 1 TeV or
below to cut off the quadratically divergent contributions.

At low energies, new physics can be integrated out and its effects are
parametrized in terms of higher dimensional operators involving only
Standard Model fields~\cite{Weinberg:1978kz}.  Precision experimental
measurements constrain the sizes of various higher dimensional
operators and consequently the scales of the corresponding new
physics~\cite{Buchmuller:1985jz}. The most stringent bounds are on the
operators  which break the (approximate) symmetries of the Standard
Model, such as those violating baryon number, flavor and CP
symmetries.  New physics which occurs at the TeV scale should respect
these Standard Model symmetries  in order not to generate any
dangerous operator with a significant size.  In the low energy
effective theory, however, there are  operators, generated by the new
physics, which conserve baryon number, flavor and CP
symmetries. Precision electroweak measurements put strong constraints
on many operators of this kind, and so far suggest no evidence for new
physics up to  $\gtrsim 5-7$ TeV~\cite{Barbieri:1999tm}. This creates
some tension with the naturalness requirement, however, which expects
new physics at $\sim$ 1 TeV to cut off the quadratic divergence to the
Higgs mass-squared. Indeed, many models which address the stabilization
of the electroweak scale have new particles in the 1 TeV range in
order to cancel the quadratic divergences incurred by the Standard
Model particles. The amount of fine-tuning required to reconcile the
difference here is not severe, and one may or may not take this
``little hierarchy problem'' seriously. Nevertheless, these
constraints definitely present an interesting challenge to theorists
trying to build models which deal with the stability of the
electroweak symmetry breaking scale.

In this paper we consider how the little hierarchy problem can be
resolved in a natural way by introducing a new symmetry at the TeV
scale. In deriving the bound of 5--7 TeV, it was assumed that  these
higher dimensional operators are generated at tree level with  ${\cal
O}(1)$ couplings to the Standard Model fields. On the other hand, the
cancellation  of the quadratic divergences involves quantum loop
diagrams only. Thus if one eliminates the tree level exchanges of the
new particles among the Standard Model fields, the bound on the scale
of the new physics can possibly be  lowered by an order of magnitude
without spoiling the cancellation of the quadratic divergences, making
the existence of the new particles in the 1 TeV range consistent with
precision electroweak data. In the next section, we propose a
symmetry, acting on the new TeV scale particles, which achieves the
above goal. In Sec.~\ref{sec:model} we present a realistic model, with
the aforementioned new symmetry, in the framework of the recently
proposed little Higgs theories,  which provide a new way to cancel the
one-loop quadratic divergences of the Higgs mass-squared and stabilize
the electroweak
scale~\cite{Arkani-Hamed:2001nc,Arkani-Hamed:2002pa,Arkani-Hamed:2002qx,Arkani-Hamed:2002qy}.
The existence of this new symmetry has many important phenomenological
consequences on future collider searches of new physics, as well as
dark matter, which will be discussed in
Sec.~\ref{sec:phenomenology}. Then we conclude in
Sec.~\ref{sec:conclusions}.

During the final stage of this project, Ref.~\cite{Wudka:2003se}
appeared which also pointed out the possibility of imposing a new
symmetry at the TeV scale to lower the scale of new physics
while evading constraints from precision measurements. The  discussion
there parallels ours in Sec.~\ref{sec:symmetry},
though the stabilization of the
electroweak scale was not addressed in that article.

\section{New symmetry for TeV scale particles}
\label{sec:symmetry}

The origin of electroweak symmetry breaking (EWSB) is one of the most
prominent questions in particle physics nowadays. If it is indeed
triggered by the vacuum expectation  value (VEV) of a scalar Higgs
doublet, naturalness arguments require new physics at or below 1 TeV
to cut off the quadratically divergent contributions to the Higgs
mass-squared. On the other hand, if EWSB is caused by some strong
dynamics, one also expects that it occurs at the  1 TeV scale in order
to obtain the EWSB scale of 246 GeV. At the Large Hadron Collider
(LHC),  the TeV scale physics will be fully explored. It is important
to be able to anticipate what kind of signals for new physics may show
up in these upcoming experiments.

Current experimental data already give some constraints on possible
new physics at the TeV scale. Absence of nucleon decays and strong
bounds on flavor-changing neutral currents indicate that these effects
cannot receive any significant contributions from the TeV scale
physics, which implies baryon number conservation and approximate
flavor symmetries at the TeV scale.  Precision electroweak
measurements also put constraints on many dimension-six operators
consistent with baryon, flavor and CP symmetries. The scales which
suppress these operators are  required to be larger than 2--7 TeV,
depending on the operators and the Higgs mass, as was discussed in
Ref.~\cite{Barbieri:1999tm}. Generally speaking these operators arise
by exchanging new heavy particles, and the bound on  the sizes of the
operators translates into the bound on the masses of the new particles
and their couplings to the Standard Model fields.  If the new
particles  are responsible for cancelling the quadratic divergences to
the Higgs mass-squared, their masses have to be at $\sim$ 1 TeV by
naturalness.  One therefore needs to worry about the compatibility of
the existence of these particles with the precision electroweak data.
Note, however, that the quadratic sensitivity to the high energy
physics of the  Higgs mass-squared parameter is a result of loop
contributions. To cancel the quadratic divergences the new particles
at the TeV  scale\footnote{The new particles can be much lighter than
1 TeV. For  simplicity we will simply call all these new particles at
or below the TeV scale TeV particles.} only need to contribute to the
Higgs mass at the loop level, {\it i.e.}, we only need interaction
vertices involving two or more new TeV particles. On the other hand,
generating higher dimensional operators at the tree level requires
different interaction vertices, those containing only one new TeV
particle.  Therefore, it is possible to suppress the tree level
contributions due to the new physics without modifying the
cancellation of the loop contributions.  The simplest and most natural
way to  implement this is to have a new symmetry acting on new TeV
particles, while all the Standard Model fields are neutral under the
new symmetry. Then there can be no interaction vertex involving the
Standard Model particles and a single new TeV particle charged under
the symmetry.  The interactions containing more than one TeV
particles, on the other hand, can still be allowed. Of course, not
every TeV scale particle would induce large higher dimensional
operators which affect the  precision electroweak measurements, so in
practice we only need the dangerous particles, for example $W'$ and
$Z'$, to be charged under this symmetry.  The simplest choice for the
new symmetry is just a $Z_2$ parity, but larger symmetry groups are
also possible. With the new symmetry, higher dimensional operators are
generated only at the loop level, and new particles as light as a  few
hundred GeV can be perfectly consistent with the precision electroweak
data.

There are existing models with such symmetry acting only on the new
particles. The  most popular and well-known example is the Minimal
Supersymmetric Standard Model (MSSM) with $R$-parity conservation. In
MSSM, all Standard Model particles have  positive $R$-parity and all
superpartners have negative $R$-parity.  Superpartner loops cancel the
quadratic divergences from the Standard Model particle  loops, but in
the low energies there is no higher dimensional operator induced by
superpartners  at the tree level. For a large portion of the parameter
space, MSSM is consistent with all the precision data. This is one of
the major reasons which make the MSSM the leading candidate for
physics beyond the Standard Model. On the other hand, without
$R$-parity, there are many strong constraints on the $R$-parity
violating couplings which require them to be unnaturally
small. Although supersymmetry is aesthetically appealing, $R$-parity
is the reality check that ensures the consistency of supersymmetric
models with precision experiments.

Another closely related example is the KK-parity in the Universal
Extra Dimensions  (UEDs), where all Standard Model particles propagate
in some number of compactified extra
dimensions~\cite{Arkani-Hamed:2000hv,Appelquist:2000nn,Cheng:2002iz,Cheng:2002ab}.
The compactification breaks the translational invariance  in the extra
dimensions down to some discrete subgroup corresponding to the
geometrical symmetry of the compactified space. As a result,  the
momentum conservation in extra dimensions is reduced to the KK-parity
conservation of the Kaluza-Klein (KK) states of the Standard Model
fields. The KK-parity prohibits the lowest  KK states from
contributing to the higher dimensional operators at the tree level,
therefore allowing them to be as light as  300
GeV~\cite{Appelquist:2000nn,Appelquist:2002wb}. The contributions from
higher KK states may also be suppressed if the mixing with the zero
mode  is small. Although the simplest UED scenario, where the KK state
loops do not cancel the quadratic divergence of the Higgs
mass-squared, does not directly address the little hierarchy problem,
the KK-parity allows the sizes of the extra dimensions to be large
enough to be probed in the near future. This feature makes the UED
model very interesting phenomenologically. In contrast,
extra-dimensional models without KK parity have much stronger bounds
on the masses of the KK states, and hence the sizes of the extra
dimensions~\cite{Cheung:2001mq}, which makes these models beyond
direct probe of near future   experiments.

The above discussion suggests that this new TeV symmetry 
could be a key to the phenomenological success
of a model with 
new particles at or below 1 TeV scale, especially one
concerning the stabilization of the electroweak scale. Recently
a new class of theories, inspired by the dimensional
deconstruction~\cite{Arkani-Hamed:2001ca,Hill:2000mu} 
and dubbed little Higgs theories,
was proposed to address the stability of the electroweak scale in a new
way~\cite{Arkani-Hamed:2001nc,Arkani-Hamed:2002pa,Arkani-Hamed:2002qx,Arkani-Hamed:2002qy}. In these theories, the quadratic 
divergence of the Standard Model loops are cancelled at one loop by new
states, with the same spin as the Standard Model particles, appearing at  
the TeV scale. The cut-off scale of the little Higgs theories can be as high as
10 TeV or above while at the same time stabilizing the electroweak
scale without fine-tuning. There are a number of variations of the little
Higgs
model~\cite{Gregoire:2002ra,Low:2002ws,Kaplan:2003uc,Chang:2003un,Chang:2003zn,Skiba:2003yf},
but in all cases so far the new TeV particles couple directly to the Standard
Model particles and one needs to worry about the 
impact on the precision electroweak physics from these new
particles.
In the next section we will show that it is possible to construct a little
Higgs model with a new parity at the TeV scale such that all the
Standard Model particles are neutral under the new symmetry. This model
therefore solves the little hierarchy problem naturally and is in good
agreement with the precision electroweak measurements.

\section{A little Higgs model}

\label{sec:model}

Little Higgs theories provide a new way to stabilize the electroweak scale.
They revive an old idea of the Higgs being a pseudo-Nambu-Goldstone 
boson (PNGB)~\cite{Georgi:yw,Georgi:1975tz,Kaplan:1983fs,Kaplan:1983sm,Georgi:1984af,Georgi:ef,Dugan:1984hq}. 
A such model is based on a chiral Lagrangian in which 
a global symmetry is both spontaneously broken and explicitly broken
by some weakly-interacting couplings. The crucial new ingredient for little
Higgs model is 
that each coupling preserves a subset of the global symmetry
under which the Higgs doublet (little Higgs) is an exact Nambu-Goldstone
boson. The little Higgs only learns its PNGB nature in the presence of more
than one set of couplings. Therefore, there is no one-loop quadratic
divergence to the little Higgs mass-squared. 
Any correction to the Higgs
mass-squared is suppressed by two loop factor relative to the cutoff,
raising the cutoff to $\sim$ 10 TeV without destabilizing the electroweak 
scale. A number of models based on various symmetry groups have been 
constructed. A universal feature is that there exist new gauge bosons, 
fermions, and scalars at the TeV scale which cancel the one-loop quadratic
divergence to the Higgs mass-squared from the Standard Model electroweak
gauge bosons, top quark, and the Higgs quartic coupling, respectively.
The corrections to the electroweak observables for several models have
been computed in
Refs.~\cite{Hewett:2002px,Csaki:2002qg,Csaki:2003si,Gregoire:2003kr}.  
In general there are strong constraints on the viable parameter space,
even though they are quite model dependent. The largest
corrections often come from the new gauge boson exchanges and VEVs of
the $SU(2)_W$ triplet scalars. It is possible to find models, with 
acceptable fine-tunings, for which such constraints are loosened in some
region of parameter space. This in turn
suggests that the tree-level corrections to the electroweak observables
are not an essential part of the little Higgs models. It is therefore
interesting to find models in which these tree-level contributions
are absent for symmetry reasons.

A natural starting point for model building is to consider moose type
models based on deconstruction. 
They often contain some geometric symmetries which may be used for
our purpose. For example, leaving out fermions for now, the minimal moose 
model in Ref.~\cite{Arkani-Hamed:2002qx}
has a reflection symmetry which exchanges the two sites if the same
$SU(2)\times U(1)$  
subgroup is gauged on each site with equal gauge coupling.
However, the way the Standard Model chiral fermions were introduced there
breaks this symmetry, and one needs a way to distribute the
Standard Model fermions evenly between the two sites.
This can be done by putting one more site in the model and placing mirror
fermions on the extra site, as will be discussed in detail
later in this section. Another important issue is that because
the non-linear sigma model is getting strongly coupled at the $\sim$ 10 TeV
scale, certain 
operators generated at that scale may be enhanced by the strong couplings,
and hence violate the bounds from the electroweak precision measurements.
In particular, the dimension-six operator involving the Higgs field,
$(h^\dagger\, D_\mu h)^2$, may be generated with 
a coefficient $\sim 1/f^2 \approx (4\pi/\Lambda)^2$, where $f \sim 1$ TeV
is the symmetry breaking scale and $\Lambda$ is the 
cutoff. This operator arises from expanding the non-linear chiral
Lagrangian,  breaks the custodial $SU(2)_C$ symmetry, and
contributes to the $\rho$ parameter. A simple way to avoid this is to
choose the global symmetry to contain an $SU(2)_C$ symmetry,
eliminating such an operator from the non-linear chiral Lagrangian.

In the following we construct a little Higgs model with a $Z_2$ symmetry
acting on the TeV scale new particles. For simplicity, 
we will call it ``$T$-parity,'' although some of the new particles may be 
lighter than 1 TeV and there may still be a few TeV particles 
even under the parity.
The $T$-parity arises due to a 
geometric reflection symmetry of our theory space, which consists of three
sites and five links. At each site $G_i$, $i=a,b,c$, there is an
$SO(5)$ global symmetry in which an $SU(2)\times U(1)$ subgroup is
gauged. The reflection symmetry ensures the gauge couplings on sites $b$ and
$c$ are equal.
The five link fields $X_j=\exp(ix_j/f)$, $j=1,\cdots, 5$, are
the non-linear sigma model fields associated with the theory 
space, as indicated in Fig.~1.
\begin{figure}[tb]
\label{figureone}
\includegraphics[width=5cm]{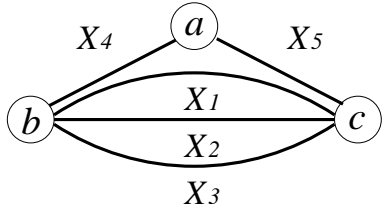}
\caption{\it The moose diagram for the theory space. It has the topology of
a torus.}
\end{figure}
This theory space is a variation  
of the minimal moose model in Ref.~\cite{Arkani-Hamed:2002qx}, with one 
additional site inserted in one of the links. However, we have chosen the
global 
symmetry to be $SO(5)$ in order to have custodial $SU(2)_C$ as an approximate
symmetry~\cite{Chang:2003un}. It has a large, approximate $[SO(5)]^{10}$
global symmetry 
spontaneously broken to $[SO(5)]^5$. The cutoff of the non-linear sigma model
is taken to be $\Lambda \sim 4\pi f \sim 10\ {\rm TeV}$. Below this cutoff
the effective theory is described by the Lagrangian
\be
\label{eq:T-op}
{\cal L} = {\cal L}_G + {\cal L}_X + {\cal L}_{\psi},
\ee
where ${\cal L}_G$ includes the kinetic terms for the $X_j$ as well as the
gauge interactions, ${\cal L}_X$ contains various plaquette operators for
the non-linear sigma model fields, and ${\cal L}_{\psi}$ involves
interactions with fermions. We now describe these three sectors in turn.

In the gauge sector, the $[SU(2)\times U(1)]^3$ gauge symmetry is
spontaneously broken to the diagonal subgroup, which is taken to
be the electroweak $SU(2)_W\times U(1)_Y$. Thus there are two sets of
heavy gauge bosons in the low energy effective theory, which can be taken
to be
\bear
A^A_\mu &\sim& (g_{b} A^{(b)}_\mu - g_{c} A^{(c)}_\mu), \quad
A^S_\mu \sim (g_{b} A^{(b)}_\mu + g_{c} A^{(c)}_\mu -2 g_{a} A^{(a)}_\mu), \\
B^A_\mu &\sim& (g_{b}^\prime B^{(b)}_\mu - g_{c}^\prime B^{(c)}_\mu), \quad
B^S_\mu \sim (g_{b}^\prime B^{(b)}_\mu + g_{c}^\prime B^{(c)}_\mu -2
g_{a}^\prime B^{(a)}_\mu), 
\eear
where $A_\mu^{(i)}, B_\mu^{(i)}$ are the $SU(2)\times U(1)$ gauge bosons at the
$i$th site and $g_{i}, g_{i}^\prime$ are the corresponding gauge couplings.  
We have chosen these particular combinations because they have definite
parity under the reflection symmetry of the theory space. $A^S_\mu$ and
$B^S_\mu$ 
are even under the interchange of sites $b$ and $c$, whereas $A^A_\mu$ and
$B^A_\mu$ are odd; this defines the $T$-parity of the heavy gauge
bosons. 
Ordinarily these heavy gauge bosons have masses of the order $g
f$, which is around $1\ {\rm TeV}$ if gauge couplings are ${\cal O}(1)$. 
However, we would like to decouple
the $T$-even heavy gauge bosons by taking the gauge group at site $a$ to be
strongly coupled: $g_{a} \sim 4\pi$. Then the $T$-even heavy gauge bosons
will be massive in the order of $10\ {\rm TeV}$, and are consisted of 
mostly site $a$ gauge bosons. This suppresses the tree-level 
contributions to the electroweak observables from the $T$-even heavy gauge
bosons, if the Standard Model fields live away from site $a$. On the other
hand, the little Higgses remain light
as they still require gauge couplings on sites $b$ and $c$, both ${\cal
O}(1)$, to know that they are not exact NGBs.

The embedding of the $SU(2)\times U(1)$ gauge group in the global $SO(5)$
is the same as in Ref.~\cite{Chang:2003un}, where the $SO(5)$ generators
are labeled as $T^l, T^r,$ and $T^v$ for the $SU(2)_l, SU(2)_r,$ and
$SO(5)/SO(4)$ generators respectively, using the $T^{la}$ generators for
$SU(2)$ and $T^{r3}$ for $U(1)$. 
It is more convenient, when we introduce fermions later, to use the 
language of $Sp(4)$, which is the universal covering group of $SO(5)$.
Throughout the paper we will use the bi-spinor notation for the 
link fields which parameterize
the coset space of $(Sp(4)\times Sp(4))/Sp(4)$. The $SU(2)_l$ and $SU(2)_r$
simply act on the upper and lower two components, respectively, of the
fundamental representation of $Sp(4)$.
In the scalar sector each link
field $X_j$ contains, under the
$SU(2)_W$ gauge group, a triplet, three singlets, and a complex doublet.
The plaquette operators we include in the Lagrangian are
\begin{eqnarray}
\label{plaqop}
{\cal L}_X &=& {\cal P}_{TS} + {\cal P}_{DS} + {\cal P}_D + 
{\cal P}_Q +{\cal P}_{T^3}
 \ , \\
{\cal P}_{TS}&=&\tau \, f^4 \, \left[ {\rm Tr}\left(\Omega X_1 X_3^\dagger
\Omega 
   X_1 X_3^\dagger \right) + {\rm Tr}\left(\Omega X_1^\dagger X_3 \Omega
   X_1^\dagger X_3\right) \right ] \nonumber \\
    &&\phantom{aa} + \tau^\prime\, f^4\, \left[
      {\rm Tr}\left(\Omega X_2 X_3^\dagger \Omega X_2 X_3^\dagger \right) +
      {\rm Tr}\left(\Omega X_2^\dagger X_3 \Omega X_2^\dagger X_3 \right) \right]
     + {\rm h.c.}\ ,\\
{\cal P}_{DS} &=&
\sigma\, 
f^4\, \left[ {\rm Tr}\left(\Omega^\prime X_4 \Omega^\prime X_4^\dagger\right)
  + {\rm Tr}\left(\Omega^\prime X_5 \Omega^\prime X_5^\dagger \right) \right]
 +{\rm h.c.}\ ,\\
{\cal P}_D &=& \omega\,f^4\, \left[ {\rm Tr}\left(\Omega X_4 \Omega X_4^\dagger\right)
  + {\rm Tr}\left(\Omega X_5 \Omega X_5^\dagger \right) \right]
 + {\rm h.c.}\ ,\\
{\cal P}_Q &=&  \lambda\,f^4\, \left[ {\rm Tr}\left( X_1 X_2^\dagger X_3
X_5^\dagger 
X_4^\dagger \right) +  {\rm Tr}\left( X_1^\dagger X_2 X_3^\dagger
X_4 X_5 \right) \right]
 + {\rm h.c.}\ , \\
{\cal P}_{T^3} &=& i \epsilon\, f^4\, {\rm Tr}\, T^{r3}
\left(X_1 X_2^\dagger X_3
X_5^\dagger X_4^\dagger + X_1^\dagger X_2 X_3^\dagger X_4 X_5
\right. \nonumber \\
&&\phantom{aaaaaaaaaaaaaaaaaaaa}\left. + X_3
X_5^\dagger X_4^\dagger X_1 X_2^\dagger + X_3^\dagger X_4 X_5 X_1^\dagger X_2
\right)    + {\rm h.c.}\ ,
\end{eqnarray}
where $\Omega = {\rm diag} (-1,-1,1,1)$ and $\Omega^\prime =
{\rm diag} (-1,-1,-1,1)$. Since we only gauge an
$SU(2)\times U(1)$ subgroup at 
each site, only two triplets and two singlets are eaten, giving masses to
the broken gauge bosons. The remaining three triplets and thirteen
singlets, as well as three doublets,
obtain masses of order $1$ TeV from plaquette operators ${\cal P}_{TS}$, 
${\cal P}_{DS}$, ${\cal P}_D$, and ${\cal P}_Q$. 
 Note that a plaquette operator of the type
$\Omega X \Omega X^\dagger$ gives mass only to scalars sitting in the
off-diagonal blocks in $X$, whereas $\Omega X \Omega X$ gives mass only
to the diagonal blocks \cite{Gregoire:2002ra}.
More specifically, the
number of scalars becoming massive through these plaquette
operators is as 
follows: two triplets and six singlets from ${\cal P}_{TS}$, two doublets
and four singlets from ${\cal P}_{DS}$, and
one doublet, one triplet and three singlets from ${\cal P}_Q$.\footnote{The
${\cal P}_D$ plaquette gives mass to the same two doublets as 
${\cal P}_{DS}$ does, and will be generated by fermion interactions
discussed later.}  
Only two electroweak doublets
remain light. Therefore in the low energies our construction gives rise to a
two Higgs doublets model. Quartic interactions of the Higgs doublets come
from the plaquette ${\cal P}_Q$, which can be analyzed using the method in
Ref.~\cite{Gregoire:2002ra}, or simply by expanding the plaquette operators
and setting all the heavy fields to zero. It is hardly surprising 
 that the Higgses
have the same quartic potential as in the $SO(5)$ minimal moose model in
Ref.~\cite{Chang:2003un}. Moreover, the $T^{r3}$ plaquettes ${\cal P}_{T^3}$
provide a Higgs mass term $ih_1 h^\dagger_2$, which is necessary to have
stable electroweak symmetry breaking, and its
coefficient $\epsilon$ is set to be a loop factor less than $\lambda$ in
${\cal P}_Q$~\cite{Chang:2003un}. All other
coefficients in ${\cal L}_X$ are ${\cal O}(1)$.

The plaquette operators in Eq.~(\ref{plaqop}) are invariant under the
reflection 
symmetry $P$ of the theory space
\bear
X_j &\leftrightarrow& X_j^\dagger, \quad j=1,2,3 \nonumber\\
X_4 &\leftrightarrow& X_5^\dagger,
\eear
and the Goldstone fields transform as
\bear
x_j &\leftrightarrow& -x_j, \quad j=1,2,3 \nonumber\\
x_4 &\leftrightarrow& -x_5.
\eear
Therefore only one linear combination of Goldstone bosons $x_4-x_5$ is even
under $P$, whose triplet and one of the singlets are eaten 
by the even gauge bosons
$A_\mu^S$ and 
$B_\mu^S$, while the rest are all odd under the reflection $P$. 
This reflection symmetry $P$ is broken once the light Higgses develop
VEVs to break the electroweak symmetry. However, there is still a
$Z_2$ parity which remains unbroken. 
To see this, we make use of the fact that,
by multiplying $\Omega$ to the link fields on both sides, a
generic Goldstone field,
\be
X = \exp(i x) = \exp i \left(
      \begin{array}{cc}
      \phi & h \\
      h^\dagger  & s
      \end{array}        \right) ,
\ee
transforms as
\be 
\Omega X \Omega = \exp (i \Omega x \Omega) = \exp i \left(
      \begin{array}{cc}
      \phi  & -h \\
       -h^\dagger &  s
      \end{array}        \right),
\ee
where the triplet $\phi$ and singlets $s$ sit in the upper left and lower
right $2\times 2$ blocks, respectively,
and the doublet $h$ sits in the off-diagonal blocks
transforming as ({\bf 2, 2}) under the $SU(2)_l\times SU(2)_r$ subgroup.
One can check
that the plaquette operators in Eq.~(\ref{plaqop}) 
are invariant under the combined operation $T=P\Omega$,
\bear
X_j &\leftrightarrow& \Omega X_j^\dagger \Omega, \quad j=1,2,3 \nonumber\\
X_4 &\leftrightarrow& \Omega X_5^\dagger \Omega.
\eear
Both light doublets, as well as the two heavy doublets, are 
even under the combined operation
$P\Omega$, which we take as the definition of $T$-parity for scalar
particles, whereas all the heavy triplets and all but two singlets are odd.
The $T$-parity remains unbroken even after the light Higgs doublets acquire
VEVs.

For the fermion sector, the reflection symmetry of the theory space forces
identical fermion contents at 
sites $b$ and $c$. Therefore we need to spread out the standard model
fermions evenly between those two sites.\footnote{The gauge group at site
$a$ is 
strongly coupled and we demand the Standard Model fermions to be
neutral under $[SU(2)\times U(1)]_a$.} We do this by introducing
additional mirror Weyl 
fermions at site $a$ and coupling them through link fields to fermions at
sites $b$ and $c$, of which a linear combination marries the mirror fermion
 to become massive in the order of 1 TeV. The orthogonal
combination remains massless and are taken to be the Standard Model
fermions. Thus it is necessary to introduce a copy of Standard Model
fermion content at each site $b$ and $c$, and a copy of mirror Standard
Model fermions at site $a$. Notice, however, that the $U(1)$ charge
assignments for all these fermions can be different from the physical Standard
Model fermions. There is some freedom in the $U(1)$
charge assignments as 
the fermions may be charged under more than one $U(1)$'s, as long as they have
the correct hypercharges under the unbroken diagonal $U(1)$.
In order to avoid large tree-level couplings between the Standard Model fermions and
the $U(1)_a$ gauge boson, we require that the fermions at site $b$ and $c$
to be neutral under $U(1)_a$. 
A convenient choice for the $U(1)$ charges, which makes the fermion mass
terms we are about to write down gauge invariant,
 is described in Table~\ref{u1charges}. Given that the $U(1)$ charge
assignments are rather
odd looking at first glance, it is quite interesting to check
that all the anomalies cancel if one includes right-handed
neutrinos.

\begin{table}[ht]
\caption{\label{u1charges} \it The $U(1)$ charge assignments for
fermions. All fermions in the table are left-handed. We denote the $SU(2)$
doublet 
quarks and leptons by $q$ and $\ell$, $SU(2)$ singlet quarks and leptons
by $u^c$, $d^c$, $e^c$, and $\nu^c$. The $U(1)$ charges for fermions 
at site $c$ are simply those of fermions at site $b$ with $U(1)_b$ 
and $U(1)_c$ charges interchanged, as required by the reflection symmetry.
The physical $U(1)_Y$ hypercharge is the sum of the $U(1)$ charges 
at all three sites.} 
\renewcommand{\arraystretch}{1.4}
\begin{ruledtabular}
\begin{tabular}[b]{c|cccccccccccc}
  &\, $\bar{q}^{(a)}\, $ &\, $q^{(b)}$\, & \,$\bar{\ell}^{(a)}$\, &\,
  $\ell^{(b)}$\, 
  &\, $\bar{u}^{c\, (a)}$ \,& \,$u^{c\, (b)}$\, &\, $\bar{d}^{c\, (a)}$\, 
  &\, $d^{c\, (b)}$\, 
  &\, $\bar{e}^{c\, (a)}\,$ &\, $e^{c\, (b)}$ &\, $\bar{\nu}^{c\, (a)}\,$
  &\, $\nu^{c\, (b)}$ \\ 
\hline
  $U(1)_a$ & 0 & 0 & 0 & 0 & $\frac12$ & 0 & $-\frac12$ & 0 & $-\frac12$ 
  & 0 & $\frac12$ & 0 \\ 
\hline
  $U(1)_b$ & $-\frac1{12}$ & $\frac1{12}$ & $\frac14$ & $-\frac14$ &
  $\frac1{12}$ & $-\frac7{12}$ & $\frac1{12}$ & $\frac5{12}$ &
  $-\frac14$ &  $\frac34$ & $-\frac14$ & $-\frac14$ \\
\hline
  $U(1)_c$ & $-\frac1{12}$ & $\frac1{12}$ & $\frac14$ & $-\frac14$ &
  $\frac1{12}$ & $-\frac1{12}$ & $\frac1{12}$ & $-\frac1{12}$ &
  $-\frac14$ &  $\frac14$ & $-\frac14$ & $\frac14$ 
\end{tabular}
\end{ruledtabular}
\end{table}

Through link fields $X_4$ and $X_5$,
the fermions at site $a$ marry with a linear combination of fermions
at sites $b$ and $c$ and become massive. To simplify notations, let us define
\begin{equation}
\tilde{X}_{i} = X_{i} + \Omega X_{i} \Omega, \quad i=4,\,5,
\end{equation}
which contain only the diagonal $2\times 2$ blocks of $X_4$ and $X_5$,
and group the fermions as follows
\begin{eqnarray}
\label{smfermion}
&& \bar{Q}^{(a)} = ( \bar{q}^{(a)}, \bar{d}^{c\,(a)}, \bar{u}^{c\,(a)}),
 \quad \quad
 \bar{L}^{(a)}= (\bar{\ell}^{(a)}, \bar{e}^{c\,(a)}, \bar{\nu}^{c\,(a)}), 
  \nonumber \\ 
&& Q^{(j)} = ( q^{(j)}, d^{c\,(j)}, u^{c\,(j)})^T,
 \quad \quad
 L^{(j)}= (\ell^{(j)}, e^{c\,(j)}, \nu^{c\,(j)} )^T, \quad j=b,\,c \, . 
\end{eqnarray}
Then, the masses of the TeV fermions can come from the interactions
\begin{equation}
{\cal L}_{fm} = \kappa_q f\, \bar{Q}^{(a)} \left( \tilde{X}_4^\dagger Q^{(b)} +
  \tilde{X}_5  Q^{(c)}  \right) 
  + \kappa_\ell f\, \bar{L}^{(a)} \left( \tilde{X}_4^\dagger L^{(b)} +
  \tilde{X}_5 L^{(c)} 
 \right).
\label{fermionmass}
\end{equation}
Given the charge assignments in Table~\ref{u1charges}, 
Eq.~(\ref{fermionmass}) is 
invariant under gauge transformations at every site. At first order in $f$,
${\cal L}_{fm}$ gives rise to a Dirac mass term
\begin{equation}
\label{tevmassive}
 \kappa_q f \, \bar{Q}^{(a)} \left( Q ^{(b)}+ Q^{(c)} \right) +
\kappa_\ell f\, 
\bar{L}^{(a)} \left(  L^{(b)} + L^{(c)} \right).
\end{equation}
These heavy fermions have masses at around 1 TeV. We shall assume that 
their masses are approximately flavor universal by some flavor symmetry 
so that they do not induce large flavor changing effects. 
The massless linear combinations 
$(Q^{(b)}-  Q^{(c)})/\sqrt{2}$ and $(L^{(b)}- L^{(c)})/\sqrt{2}$ become
the Standard Model fermions in the low energies. The 
interactions in Eq.~(\ref{fermionmass}) also induce the
plaquette operators 
${\cal P}_{D}$, which lift extra scalar doublets, 
through loops.\footnote{${\cal P}_{DS}$ can also be generated radiatively if
we do not include right-handed neutrinos or they have different couplings
from charged leptons.}

For Yukawa couplings, we first concentrate on the top sector, which must be
dealt with in a way without introducing quadratic divergences to the Higgs
mass. Toward this end we introduce additional vector-like colored fermions 
$\psi_{u,\,d}^{(j)},\, \psi_{u,\,d}^{c\,(j)},\, j=b,\,c$, with 
$\psi_{u}^{(j)}(\psi_{d}^{(j)})$ having the opposite $U(1)$ charges of 
$u^{c\,(j)}(d^{c\,(j)})$.
Defining
\begin{equation}
{\cal Q}^{(j)} = (q_3^{(j)}, \psi_u^{(j)}, \psi_d^{(j)})^T, \quad \quad
{\cal U}^{c\, (j)}=(0_2, u_3^{c\, (j)}, 0), \quad j = b, c \quad,
\label{topfermion}
\end{equation}
the top Yukawa coupling is then generated by 
\begin{eqnarray}
{\cal L}_{\rm top} &=& y_1 f \left( {\cal U}^{c\,(b)} {X}_1
{X}_3^\dagger {\cal Q}^{(b)} + {\cal U}^{c\,(c)}\, {\Omega}
{X}_1^\dagger 
{X}_3 {\Omega}\, {\cal Q}^{(c)} \right)  \nonumber\\
&& + f \sum_{j=b,c}\left( y_2\, \psi_u^{(j)}\psi_u^{c\,(j)} 
   + y_3\, \psi_d^{(j)}\psi_d^{c\,(j)} \right) .
\label{topyukawa}
\end{eqnarray}
There is only one massless component left at this stage, which we take to
be the physical top quark,
\begin{equation}
t = \frac1{\sqrt{2(y_1^2+y_2^2)}} \left[ y_1 (u_3^{(b)}-u_3^{(c)})
  + y_2 ( \psi_u^{(b)} - \psi_u^{(c)} ) \right] ,
\end{equation}
and the top Yukawa coupling turns out to be
\begin{equation}
 \lambda_t \sim \frac{y_1 y_2}{\sqrt{y_1^2 + y_2^2}} .
\end{equation}
For the other Yukawa couplings we can simply write down operators in the
first line in Eq.~(\ref{topyukawa}) without introducing additional vector-like
fermions, $\psi_u$ and $\psi_d$, in Eq.~(\ref{topfermion}). These other
Yukawa couplings introduce one-loop quadratic divergences without
destabilizing the electroweak scale, since the divergences are suppressed
by the smallness of the Yukawa couplings.

The $T$-parity of the fermion is defined as $T= (-1)^F P$, where $F$ counts
the fermion number and $P$ is the reflection that interchanges sites
$b$ and $c$. The reason for the extra minus sign from $(-1)^F$ is because
fermions at site $a$, even under the reflection $P$, should be
$T$-odd since it becomes heavy through the Dirac mass term
Eq.~(\ref{tevmassive}). This minus sign in turn gives even $T$-parity for
the Standard Model fermions, as desired. 
Together with $T= {\Omega} P$ for the scalars, the
interactions in ${\cal L}_{fm}$ and ${\cal L}_{\rm top}$ are invariant
under the $T$-parity, which explains the insertion of the ${\Omega}$
operators in those interactions. Moreover, the linear combination that
becomes massive 
in the TeV range in Eq.~(\ref{tevmassive}) is odd, whereas the massless
combination, which becomes the Standard Model fermion in the low energies,
is even. All the heavy fermions, except for the following two 
combinations\footnote{Note that the mass of the $\psi_d$ 
field can be lifted to 
$4\pi f \sim 10\ {\rm TeV}$ without spoiling naturalness, due to an
accidental $SU(3)$ symmetry discussed in Ref.~\cite{Chang:2003un}.} 
\begin{eqnarray}
t^\prime &=& \frac1{\sqrt{2(y_1^2+y_2^2)}} \left[ y_2 (u_3^{(b)}-u_3^{(c)})
  - y_1 ( \psi_u^{(b)} - \psi_u^{(c)} ) \right] , \\
\tilde{d}^\prime &=& \frac1{\sqrt{2}} \left( \psi_d^{(b)} -
\psi_d^{(c)} \right),
\end{eqnarray}
are odd under the $T$-parity. To summarize, we define the $T$-parity as
\begin{eqnarray}
T &=& P,\phantom{\Omega (-1)^F} \quad \quad {\rm for\ vector\ bosons,} \\
T &=& \Omega P,\phantom{(-1)^F} \quad \quad {\rm for\ scalars,} \\
T &=& (-1)^F P,\phantom{\Omega} \quad \quad {\rm for\ fermions,}
\end{eqnarray} 
and it is a symmetry of our Lagrangian and remains unbroken
after electroweak symmetry breaking. 

As mentioned earlier, our theory space is a modification of the two-site,
minimal $SO(5)$ moose model, whose precision electroweak physics was
analyzed in 
some details in Ref.~\cite{Chang:2003un}. There it was shown that 
the most dangerous contributions are from those due to tree-level heavy gauge
boson exchanges, including modification of electroweak currents, four
fermion operators, and custodial $SU(2)_C$
breaking. The custodial $SU(2)_C$ is a good symmetry of the
non-linear chiral Lagrangian, but is spontaneously broken by the misalignment
of the two Higgs VEVs as a result of the commutator type quartic
Higgs coupling. The breaking of $SU(2)_C$ then shows up in the
TeV gauge boson couplings~\cite{Chang:2003un}.  In our model, the $SU(2)_C$
is also spontaneously broken by the Higgs VEVs, but the TeV gauge bosons
are odd 
under $T$-parity and do not contribute at tree level. The $T$-even heavy
gauge bosons are in the 10 TeV range and have small couplings to the
Standard Model fermions and Higgses, who live away from site $a$. So the 
contributions due to the $T$-even heavy gauge bosons are also very
small. The $SU(2)_W$ triplet scalars are 
all odd under $T$-parity and therefore do not obtain VEVs at all, since
the tadpole term is forbidden by $T$-parity. There are several $T$-even
scalar doublets and 
singlets at the TeV scale, but their tree-level contributions are either
suppressed 
by the small Yukawa couplings of light Standard Model fermions or, in the
case for top quark, weakly constrained due to lack of precision data. 
The leading contributions to the electroweak
observables in this model come from loops of the two Higgs doublets and
top partners. These contributions were also discussed in 
Ref.~\cite{Chang:2003un} and in general are safe for a wide range of
model parameters.

Finally, before concluding this section, one may ask since the site-$a$ gauge
bosons are very heavy, one should be able to integrate them out and obtain an
effective two-site model at low energies.\footnote{We are indebted to Nima
Arkani-Hamed for inspiring conversations on this issue.} The question is
then how the 
fermion interactions 
preserve the $Z_2$ parity in this two-site effective theory.
To this end we note that with just sites $b$, $c$ and links between them,
the object
\begin{equation}
 X^\dagger \, D_{\mu}^{(b)} X =  X^\dagger 
\left( \partial_\mu + i g_b A_{\mu}^{(b)} \right) X ,
\end{equation}
which is invariant under gauge transformations at site $b$,  
transforms in the same way as $i g_c A_\mu^{(c)}$. Therefore, we can write
down the following gauge invariant interaction for the fermion living on
site $c$,
\begin{equation}
\bar{\psi\,} i \bar{\sigma}^\mu \left[ \partial_\mu + r\, i g_c A_\mu^{(c)}
+ (1-r)  X^\dagger ( D_\mu^{(b)} X ) \right] \psi \, .
\end{equation}
For $g_b=g_c$ and $r=1/2$, the fermion just couples to the massless 
even combination of the gauge fields. The coupling of a light fermion,
a heavy odd fermion, and the odd gauge bosons can also be reproduced by the
interaction
\begin{equation}
\bar{\psi}_{\rm light} \frac{i \bar{\sigma}^\mu}{2} \left[ i g_c A_\mu^{(c)}
-  X^\dagger ( D_\mu^{(b)} X ) \right] \psi_{\rm heavy} + {\rm h.c.}
\, .
\end{equation}
The exact $Z_2$ symmetry is not transparent in this language. Nevertheless,
it may serve as a useful guide to construct other type of little Higgs models 
with the $T$-parity.

\section{Phenomenological consequences}
\label{sec:phenomenology}

To be consistent with the electroweak data, the new TeV particle symmetry
can be just an approximate symmetry. However, it is well motivated to keep
this symmetry exact. In this case, there are many interesting 
phenomenological consequences, so we will concentrate on the case of an exact
symmetry in this section.

Since all Standard Model particles are neutral under the new TeV symmetry, the
lightest particle which transforms non-trivially under this symmetry will
be stable. Here we call it the LTP, the lightest $T$-odd particle. 
This new symmetry has important implications for future collider
experiments, as new particles charged under it cannot be
singly produced. Direct searches have to rely on pair-productions. In addition,
after they are produced, they will decay to the LTP which is stable. If the
LTP is electrically charged, it will give rise to charged tracks in the
detector which are easy to identify. However, a charged LTP is not favored
as it causes cosmological problems. On the other hand, the
neutral LTP will escape the detector, resulting in missing energy signals.
Most of the collider phenomenology studies for little Higgs theories so far
do not assume this new TeV symmetry, and the Standard Model fermions can
interact directly with a single TeV gauge 
boson\cite{Burdman:2002ns,Han:2003wu,Sullivan:2003xy}. Similar to the usual
$W'$ and $Z'$ searches, these studies rely on single TeV gauge boson
productions; neither is there a new stable particle in the decay
products. Hence the existence of this new TeV symmetry escapes conclusions
from these
previous studies, except Ref.~\cite{Han:2003gf} which studies
loop induced processes, and dramatically 
alters the collider phenomenology.

In fact, the collider phenomenology with this new TeV symmetry is similar
to that of the $R$-parity conserving supersymmetric theories 
(and KK-parity conserving
UEDs). The typical signals are jets and/or leptons plus missing energies from
decays of heavy new particles with odd $T$-parity. To 
distinguish it from supersymmetry, we need to know the spins of these
new particles which probably requires a lepton collider or extra 
efforts for a hadron collider. Unlike supersymmetric models, 
new particles in a little Higgs theory have the 
same spins as the corresponding Standard Model particles whose quadratic
divergences they are supposed to cancel. 
On the other hand, in UEDs the KK excitations 
also have the same spins as the Standard Model particles.
The difference between our little Higgs model
and UEDs is that there is no reason for all the new TeV particles to be
closely degenerate in our model. Thus the jets and leptons from decays 
of the TeV particles in general will not be soft in the little 
Higgs model, unlike in the UEDs, which makes their detection easier.
Moreover, we do not expect to see the second KK level states at energies
not far above these TeV particles.

The existence of a stable weakly-interacting neutral particle, like the LTP,
has important astrophysical implications. In this regard the LTP shares 
properties similar to the
the lightest supersymmetric particle (LSP) in $R$-parity conserving
supersymmetric Standard Model and the lightest KK-particle (LKP) in UED
models. It can be a good dark matter candidate if
it is neutral under the unbroken Standard Model gauge group. 
In the little Higgs model 
we proposed, the best candidates are the 
$B'$ gauge boson and the $SU(2)_W$ singlets and neutral components of triplets
in the scalar
link fields. For $B'$ LTP, it is similar to the case studied in the
UEDs~\cite{Cheng:2002iz,Cheng:2002ab}. 
The $B'$ LTP gives the right relic density for dark matter if its
mass is in the range of 600 GeV -- 1.2 TeV~\cite{Servant:2002aq}, 
which is consistent
with the little Higgs model. The detection rates of $B'$ LTP in various
dark matter detection experiments are quite different from those of the LSP in
the supersymmetric
theories~\cite{Cheng:2002ej,Hooper:2002gs,Servant:2002hb,Bertone:2002ms,Majumdar:2002mw}.
In particular, because the annihilation
of two $B'$s into Standard Model fermions are not chirally suppressed, the
indirect  
detection of $B'$ LTPs annihilating into electron-positrons, neutrinos,
and photons are much more promising than those of the LSP. For example, a
peak in the positron 
energy distribution at the mass of $B'$ may be seen in AMS, the anti-matter
detector to be placed on the International Space Station, which is
nonetheless not the case for the LSP~\cite{Cheng:2002ej}.

As for scalar dark matter, it was recently studied in
Ref.~\cite{Birkedal-Hansen:2003mp} for a different little Higgs model. 
In that model
there is also an exact discrete symmetry, except that the heavy gauge
bosons are neutral under that symmetry and the electroweak constraints are
still a concern.
It was found that there are two mass ranges for which the scalar LTP can
give rise to the right relic density for dark 
matter: a low mass $\sim$ 100 GeV if the LTP is mostly an $SU(2)_W$ singlet,
and a high mass range $\gtrsim$ 500 GeV if it is a mixture of the 
singlet and the neutral component of the 
triplet. In our case, the singlet annihilates through neither the weak gauge
bosons nor the TeV gauge bosons, contrary to the case in
Ref.~\cite{Birkedal-Hansen:2003mp}, so it has to be even lighter than the
low mass region in order
to obtain the right relic density. On the other hand, the $SU(2)_W$ triplet
scalars do interact with the light gauge bosons, and hence the estimate 
in Ref.~\cite{Birkedal-Hansen:2003mp} for the high mass range should roughly
apply.

\section{Conclusions}
\label{sec:conclusions}
We have shown that, by proposing a symmetry acting only on new particles in
the TeV scale, it is possible to relax the constraints, coming from the
precision electroweak measurements, on the scale of new physics, thereby
resolving the little hierarchy problem. The critical observation is that
these constraints can be lowered by an order of magnitude if there is no
tree-level exchanges of new heavy particles, which require interaction
vertices containing only one new particle, among the Standard Model
particles. On the other hand, stabilization of the electroweak scale
necessitates cancellations of quadratic divergences to the Higgs
mass-squared, which involve quantum loop diagrams and entail vertices with
more than one new particles. Thus if one imposes a symmetry to eliminate
the tree level exchanges of the heavy states, the electroweak scale can be
stabilized naturally without conflicting with the precision measurements.

There are existing models with this kind of new symmetry in the TeV scale,
for example the $R$-parity for supersymmetric theories and the KK-parity
for UEDs.
In this paper, we present a new model, in an entirely different class,
with such a new TeV 
symmetry. It is a little Higgs model implemented with a $Z_2$ symmetry in
the TeV 
scale which we call $T$-parity. This model has new particles at around 1
TeV, which are responsible for cutting off one-loop quadratic divergences,
due to the 
Standard Model particles of the same spin, to the Higgs mass-squared and
stabilizing the electroweak scale up to 10 TeV
without fine-tuning. At the same time, it is compatible with the
precision data, a nice consequence of the $T$-parity. It is also an
intriguing observation that all anomalies cancel in our model with
the somewhat sophisticated $U(1)$ charge assignments.

The existence of this new TeV symmetry has many important implications for
phenomenology, in addition to solving the little hierarchy problem. The
lightest new particle charged under this TeV symmetry, the
LTP, is a weakly-interacting stable particle. The LTP serves as a good
candidate for dark matter if it is also neutral under the Standard Model
gauge group, a property similar to its counterparts, the LSP in
supersymmetric theories with $R$-parity and the LKP in UEDs. In terms of
collider phenomenology,  the typical signals are 
jets and/or leptons plus missing energies due to decays of $T$-odd heavy
particles, which feature is shared by all three classes of theories: the
supersymmetric theories with $R$-parity, UEDs with KK-parity, and little
Higgs theories with $T$-parity. Nevertheless, more detailed studies
can potentially tell these three categories of theories apart from one another.

As the completion date of the LHC approaches, the mystery of the TeV scale
physics will be unraveled in the near future. Obviously it is of great
interest to further explore the consequences of this new TeV symmetry in
finer detail, whether they are specific to models or generic to 
theories with the new symmetry. 

\begin{acknowledgments}
We thank N.~Arkani-Hamed, P.~Creminelli, M.~Schmaltz, N.~Toumbas, and J.~Wacker
for useful discussions. We also thank the Aspen Center for Physics
for hospitality where part of this work was completed.
This work is supported in part by the National Science Foundation under
grant PHY-98-02709.
\end{acknowledgments}


\end{document}